\documentclass[superscriptaddress,twocolumn,bibnotes,
amsmath,amssymb,aps,pra,floatfix]{revtex4-2} 

\usepackage{graphicx}% Include figure files
\usepackage{dcolumn}% Align table columns on decimal point
\usepackage{bm}% bold math
\usepackage{hyperref}% add hypertext capabilities
\usepackage{mathtools}
\usepackage{mathrsfs}

\usepackage{braket}
\usepackage{siunitx} %this allows \SI{}{} specifications of units
\usepackage{multirow} %for tables
\newcommand{\be}{\begin{equation}} 
\newcommand{\ee}{\end{equation}} 
\DeclareMathOperator{\sech}{sech}

\newcommand{\JILA}{JILA, NIST and Department of Physics, University of Colorado, Boulder, USA}
\newcommand{\CTQM}{Center for Theory of Quantum Matter, University of Colorado, Boulder, CO 80309, USA}
\newcommand{\Toronto}{Department of Physics and CQIQC, University of Toronto, Ontario M5S~1A7, Canada}

\begin{document}

\title{\texorpdfstring{Observation of unitary  p-wave interactions\\ between fermions in an optical lattice}{}}

\author{V.\ Venu}
\altaffiliation{These authors contributed equally to the work.}
\affiliation{\Toronto}
\author{P.\ Xu}
\altaffiliation{These authors contributed equally to the work.}
\affiliation{\Toronto}
\author{M.\ Mamaev}
\affiliation{\JILA}
\affiliation{\CTQM}
\author{F.\ Corapi}
\affiliation{\Toronto}
\author{T.\ Bilitewski}
\affiliation{\JILA}
\affiliation{\CTQM}
\author{J.~P.~D'Incao}
\affiliation{\JILA}
\author{C.\ J.\ Fujiwara}
\email{cora.fujiwara@utoronto.ca}
\affiliation{\Toronto}
\author{A.\ M.\ Rey}
\email{arey@jilau1.colorado.edu}
\affiliation{\JILA}
\affiliation{\CTQM}
\author{J.\ H.\ Thywissen}
\email{joseph.thywissen@utoronto.ca}
\affiliation{\Toronto}

\date{\today}

\begin{abstract} 
Exchange-antisymmetric pair wavefunctions in fermionic systems can give rise to unconventional superconductors and superfluids with non-trivial transport properties \cite{Huebener,Volovik,Ivanov2001,Mizushima2016}. The realisation of these states in controllable quantum systems, such as ultracold gases, could enable new types of quantum simulations \cite{Botelho2005,Gurarie2005,Cheng2005,Levinsen2007PRL}, topological quantum gates \cite{Tewari2007,Zhang2007,Nayak2008}, and exotic few-body states \cite{Lasinio2008PRA,DIncao2008PRA,Nishida2013PRL,Wang2011PRL}. However, p-wave and other antisymmetric interactions are weak in naturally occurring systems \cite{Martin2013,Lemke2011,top2021}, and their enhancement via Feshbach resonances in ultracold systems \cite{Regal:2003go,Zhang:2004cy} has been limited by three-body loss \cite{Suno2003PRL,Schunck:2005cf,Gunter2005,chevy2005,Gaebler:2007,Inada:2008hz}. In this work, we create isolated pairs of spin-polarised fermionic atoms in a multi-orbital three-dimensional optical lattice. We spectroscopically measure elastic p-wave interaction energies of strongly interacting pairs of atoms near a magnetic Feshbach resonance, and find pair lifetimes to be up to fifty times larger than in free space. We demonstrate that on-site interaction strengths can be widely tuned by the magnetic field and confinement strength, but collapse onto a universal single-parameter curve when rescaled by the harmonic energy and length scales of a single lattice site. Since three-body processes are absent within our approach, we are able to observe elastic unitary p-wave interactions for the first time. We take the first steps towards coherent temporal control via Rabi oscillations between free-atom and interacting-pair states. All experimental observations are compared both to an exact solution for two harmonically confined atoms interacting via a p-wave pseudopotential, and to numerical solutions using an ab-initio interaction potential. The understanding and control of on-site p-wave interactions provides a necessary component for the assembly of multi-orbital lattice models \cite{pbandreviewLi,Dutta2015}, and a starting point for investigations of how to protect such a system from three-body recombination even in the presence of weak tunnelling, for instance using Pauli blocking and lattice engineering. This combination will open a path for the exploration of new states of matter and many-body phenomena enabled by elastic p-wave interactions \cite{Botelho2005,Gurarie2005,Cheng2005,Nayak2008,Ivanov2001}.
\end{abstract}
\maketitle

The emergent behaviour of a quantum many-body system is fundamentally tied to the quantum statistics of its constituents. For pairs of identical fermions, the wavefunction must be exchange antisymmetric, which is found only in odd-$L$ pairwise collision channels, where $L$ is orbital angular momentum. Despite a well understood connection between odd-$L$ interactions and topological properties \cite{Ivanov2001,Botelho2005,Gurarie2005,Cheng2005,Tewari2007,Nayak2008,Alicea2012}, liquid $^3$He remains the only laboratory example of well established p-wave ($L=1$) interactions. The discovery of tunable p-wave interactions in ultracold atoms \cite{Regal:2003go,Zhang:2004cy} was promising, but experimental efforts have so far been severely limited by enhanced three-body recombination, a process where three atoms collide to form a diatomic molecule, releasing enough kinetic energy to make all products escape confinement \cite{Suno2003PRL,Schunck:2005cf,Gunter2005,chevy2005,Gaebler:2007,Inada:2008hz}. The essential challenge for $L>0$ systems is that wavefunction amplitude at short inter-nuclear separation, where recombination processes are strong, is enhanced by centrifugal kinetics. Progress has been made in understanding few-body correlations \cite{Levinsen2007PRL,Lasinio2008PRA,DIncao2008PRA,Nishida2013PRL,Wang2011PRL} and developing proposals towards overcoming this obstacle via wavefunction engineering \cite{han2009PRL}, including low-dimensional confinement \cite{Chang1Dloss,marcum1Dloss}. Still, p-wave interaction energies between free atoms have yet to be measured directly or compared to predictions of any theory. Even at the level of two particles, the description of p-wave interactions by a Feshbach-tuned, energy-dependent scattering volume $v(\mathcal{E})$ \cite{Idziaszek,Blume04}, has yet to be tested experimentally.

In this article, we report the first direct measurement and coherent control of the elastic p-wave interaction between two identical fermions in a multi-orbital lattice. Central to this advance is the use of strong three-dimensional (3D) confinement to modify the wavefunction and to suppress three-body processes. Interactions are tuned using the magnetic Feshbach coupling  \cite{Cheng2005} between free-atom pairs and a molecular dimer channel. Our spectral resolution and orbital control allow us to transfer pairs of weakly interacting $^{40}$K atoms into strongly interacting two-atom complexes whose energies and wavefunctions separate them into repulsive and attractive  branches. Within the two lowest branches we are able to reach the unitary limit, where $v(\mathcal{E})$ diverges. We demonstrate the coherence of the conversion process between non-interacting and strongly interacting atomic pairs by measuring Rabi oscillations between them, and find an oscillation frequency consistent with theory. Finally, we demonstrate that losses in the upper branch are limited by the intrinsic lifetime of the $^{40}$K molecular dimer, and we observe lifetimes that are fifty times larger than observed previously for weakly confined p-wave dimers of $^{40}$K.

\begin{figure*}[!htb]
\centering
\includegraphics[width=1.0\linewidth]{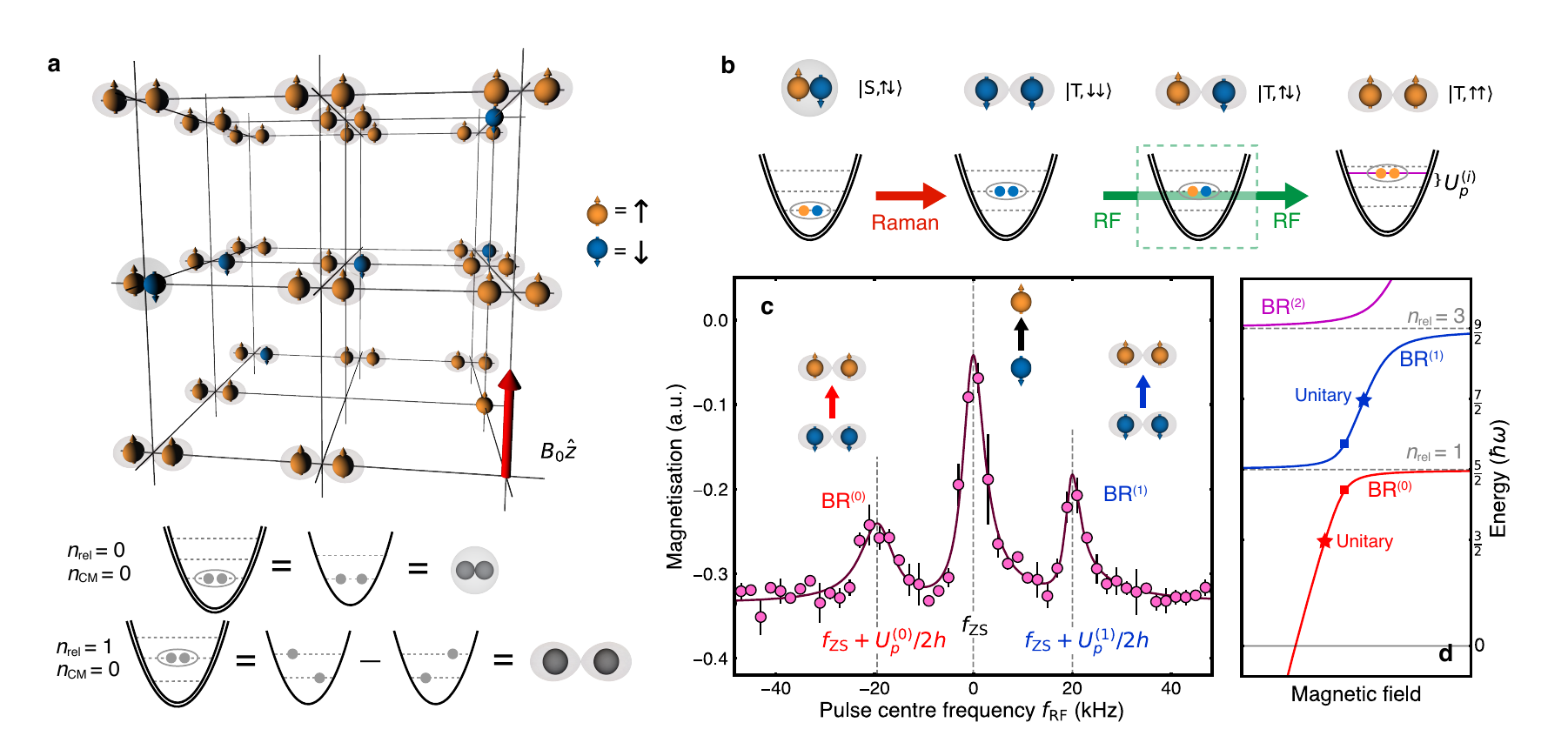} 
\caption{{\bf Spectroscopy of p-wave interactions between spin-polarised fermions.} {\bf a}, Atoms with $\uparrow$ or $\downarrow$ spin are loaded into a harmonic trap array formed by a deep 3D optical lattice. The double-struck parabola represents the two-atom $n_\mathrm{rel}$ motional degree of freedom. Pairs of atoms are shown in the non-interacting ground ($n_{\mathrm{rel}}=n_{\mathrm{CM}}=0$) or first excited ($n_{\mathrm{rel}}=1$, $n_{\mathrm{CM}}=0$) motional mode. An applied magnetic field ($B_0\hat{z}$) creates the Feshbach coupling between $\uparrow$ atoms.
{\bf b}, Measurement protocol. An optical Raman $\pi$ pulse converts singlet $\ket{\mathrm{S},\uparrow\downarrow}$ pairs in the ground motional state into $\ket{\mathrm{T},\downarrow\downarrow}$ with a motional excitation $n_{\mathrm{rel}}=1$. A radio-frequency sweep, centred at $f_\mathrm{RF}$, then transfers some pairs to the interacting $\ket{\mathrm{T},\uparrow\uparrow}$ state through a two-RF-photon process. The off-resonant intermediate state $\ket{\mathrm{T},\uparrow\downarrow}$ is shown in the dashed box.
{\bf c}, The measured magnetisation (here, at magnetic field 200.00(1)\,G and  trap  angular frequency $\omega=2\pi \times 129(2)\,\mathrm{kHz}$) exhibits three distinct spectroscopic features for varying RF centre frequency. The left-most and right-most peaks correspond to transitions to interacting states in the $\mathrm{BR}^{(0)}$ and $\mathrm{BR}^{(1)}$ branches, with interaction energies $U_p^{(0)}$ and $U_p^{(1)}$, as labelled. The solid line is a best-fit spectral function. 
{\bf d}, The spectrum of interacting $\ket{T,\uparrow\uparrow}$ pairs reflects mixing of the odd-$n_{\mathrm{rel}}$ harmonic states with a magnetic dimer state. Squares correspond to the spectral peak locations from {\bf c}, and stars indicate points with unitary p-wave interactions.
}
\label{fig:intro}
\end{figure*}

Our optical lattice system realises an array of isotropic harmonic traps, each occupied by a pair of atoms with spin and orbital degrees of freedom (Fig.~\ref{fig:intro}a). The spin state of a pair is $\ket{\mathcal{S}, M_{\mathcal{S}}}$, where $\mathcal{S} = \{\mathrm{S},\mathrm{T}\}$ indicates either singlet or triplet spin symmetry, $M_{\mathcal{S}} = \{\uparrow\uparrow, \downarrow\downarrow,\uparrow\downarrow\}$ are projections on the magnetic field axis, and $\uparrow$ and $\downarrow$ are the lowest hyperfine states of $^{40}$K. Tunable enhancement of p-wave interactions is provided by a Feshbach resonance for spin-symmetric pairs $\ket{\mathrm{T},\uparrow\uparrow}$ (Methods). The motional state of a pair is described by the relative and centre-of-mass mode numbers $n_{\mathrm{rel}}=\left\{0,1,\dots \right\}$ and $n_{\mathrm{CM}}=\left\{0,1,\dots \right\}$ respectively. The centre of mass decouples from the collisional interactions and remains in its motional ground state,  $n_{\mathrm{CM}}=0$. The relative mode number is $n_{\mathrm{rel}}=2\mathcal{N} + L$, where $\mathcal{N}$ is the conventional radial excitation number for a spherical harmonic oscillator. Since the overall pair state must have odd exchange symmetry and the interacting spin state $\ket{\mathrm{T},\uparrow\uparrow}$ is even, the motional state must have odd $L$, which implies $n_{\mathrm{rel}}=1,3,\dots$ for $L=1$ (p-wave). This is in contrast to s-wave-interacting spin singlet states which can interact when prepared in the least energetic motional mode ($n_{\mathrm{rel}}=n_{\mathrm{CM}}=0$) \cite{esslingerswave,Hartke2022}.

The magnetic-field-dependent eigenstates of a $\ket{\mathrm{T},\uparrow\uparrow}$ pair can be understood as the coupling of the odd-$n_{\mathrm{rel}}$ motional modes to a molecular state. We sketch the spectrum of the interacting pair in Fig.~\ref{fig:intro}d. For fields far below the Feshbach resonance, the spectrum is given by a ladder with harmonic spacing $2\hbar\omega$ (corresponding to $n_{\mathrm{rel}}=1,3,\dots$), where $\omega$ is the trap angular frequency, and a molecular dimer state whose energy depends linearly on magnetic field. As each motional mode becomes near resonant with this dimer, the Feshbach coupling imparts a p-wave interaction energy shift and mixes the harmonic states. We label the resulting eigenstates of the interacting pair as branches $\{\mathrm{BR}^{(0)},\mathrm{BR}^{(1)},\mathrm{BR}^{(2)}\dots \}$ in order of increasing energy $\mathcal{E}^{(i)}$. In this work, we probe the lowest energy branches, $\mathrm{BR}^{(0)}$ and $\mathrm{BR}^{(1)}$. Since they are both adiabatically connected to the $n_{\mathrm{rel}}=1$  mode, we use it as a common reference to define the on-site interaction energies $U_{p}^{(i)}$, i.e.,
$\mathcal{E}^{(0)} = U_{p}^{(0)}+\frac{5}{2}\hbar\omega$ and $\mathcal{E}^{(1)} = U_{p}^{(1)}+\frac{5}{2}\hbar\omega$. 

\begin{figure*}[!htb]
\centering
\includegraphics[width=1.0\linewidth]{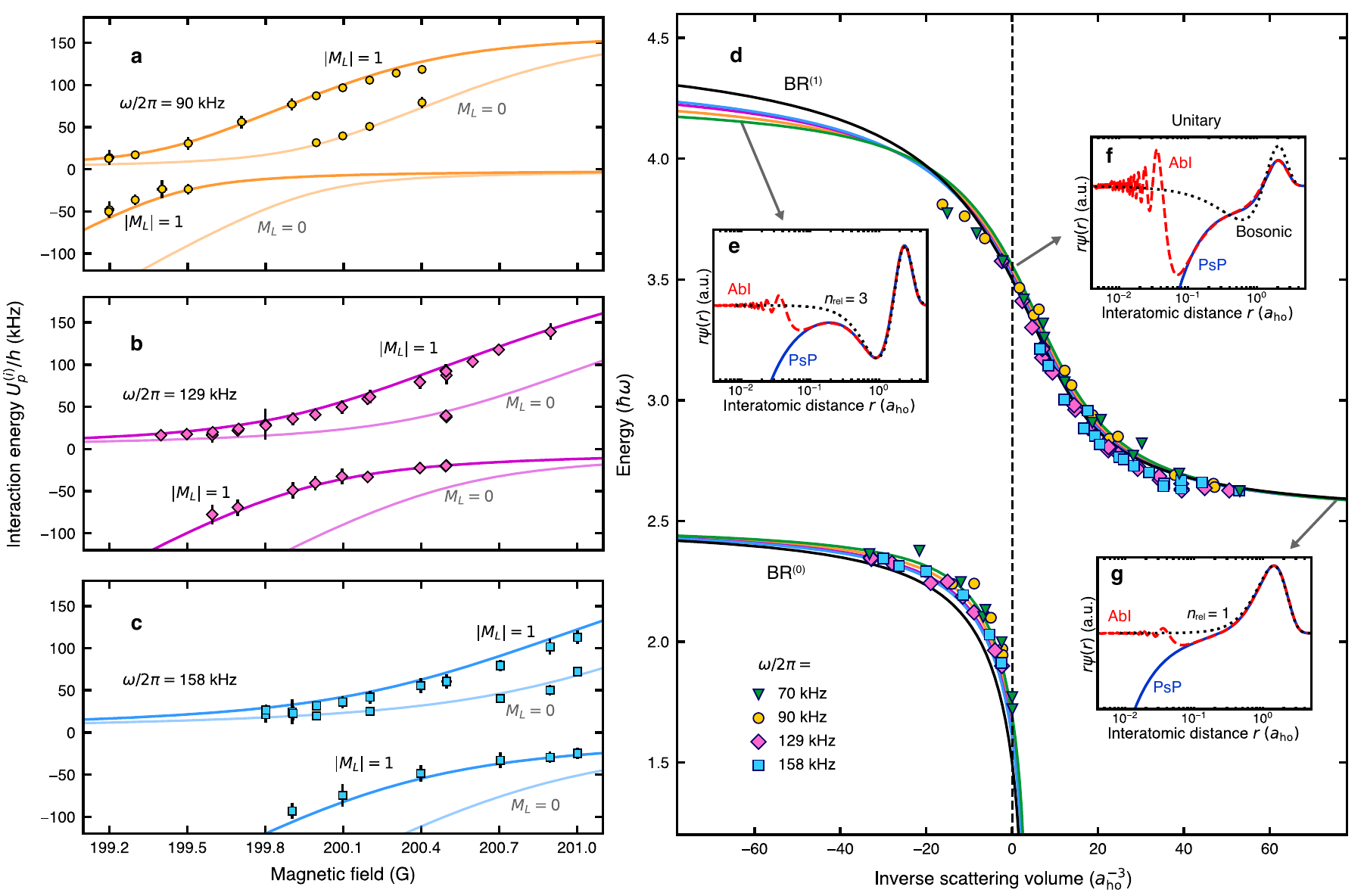}   
\caption{{\bf Characterisation of unitary and elastic p-wave interactions.} {\bf a}-{\bf c}, The measured interaction energies (points) are shown versus magnetic field at three different trap frequencies. Vertical error bars are FWHM values from the best-fit spectral function. Solid lines are the pseudopotential predictions for angular momentum projections $M_L=-1,0,1$, including anharmonic corrections (Methods). {\bf d}, When scaled by the harmonic oscillator angular frequency, the measured energies collapse onto a single universal curve as a function of inverse scattering volume, in units of the oscillator length $a_{\mathrm{ho}}$. The black solid line is the harmonic pseudopotential energy, and the coloured solid lines include anharmonic corrections. 
{\bf e}-{\bf g}, Spatial wavefunctions of the interacting pairs for various scattering volumes. The solid blue lines are obtained from the pseudopotential (PsP), and the red dashed lines from an ab-initio (AbI) calculation (see Methods). The short-range divergence of the PsP wavefunction requires a cutoff in order to be normalisable. Since {\bf e} and {\bf g} are in the non-interacting limit with $n_{\mathrm{rel}}=3$ and $n_{\mathrm{rel}}=1$ motional quanta respectively, the corresponding oscillator states are shown as black dotted lines. Panel {\bf f} corresponds to the unitary limit, where for comparison, a non-interacting bosonic pair wavefunction in the $n_\mathrm{rel}=2$ state is shown with a black dotted trace.  
}
\label{fig:energy}
\end{figure*}

We assemble the desired pair states by orbital excitation of a low-entropy spin mixture. First, $\ket{\mathrm{S},\uparrow\downarrow}$ pairs in the lowest motional mode $(n_{\mathrm{rel}}=n_{\mathrm{CM}}=0)$ are created by loading a spin-balanced degenerate Fermi gas into a 3D optical lattice of moderate depth (Methods). The lattice depth is then rapidly increased, which isolates atom pairs and prevents undesired three-body processes. Orbital excitation and triplet spin symmetry is created by a $\pi$-pulse from optical Raman beams, whose detuning from the electronic excited state is chosen to minimise photoassociative loss of pairs (Methods). The pulse transforms $\ket{\mathrm{S},\uparrow\downarrow}$ pairs into the spin-symmetric state $\ket{\mathrm{T},\downarrow\downarrow}$ with a relative orbital excitation $n_{\mathrm{rel}}=1$ (see Fig.~\ref{fig:intro}b).

\begin{figure*}[!htb]
\centering
\includegraphics[width=1\linewidth]{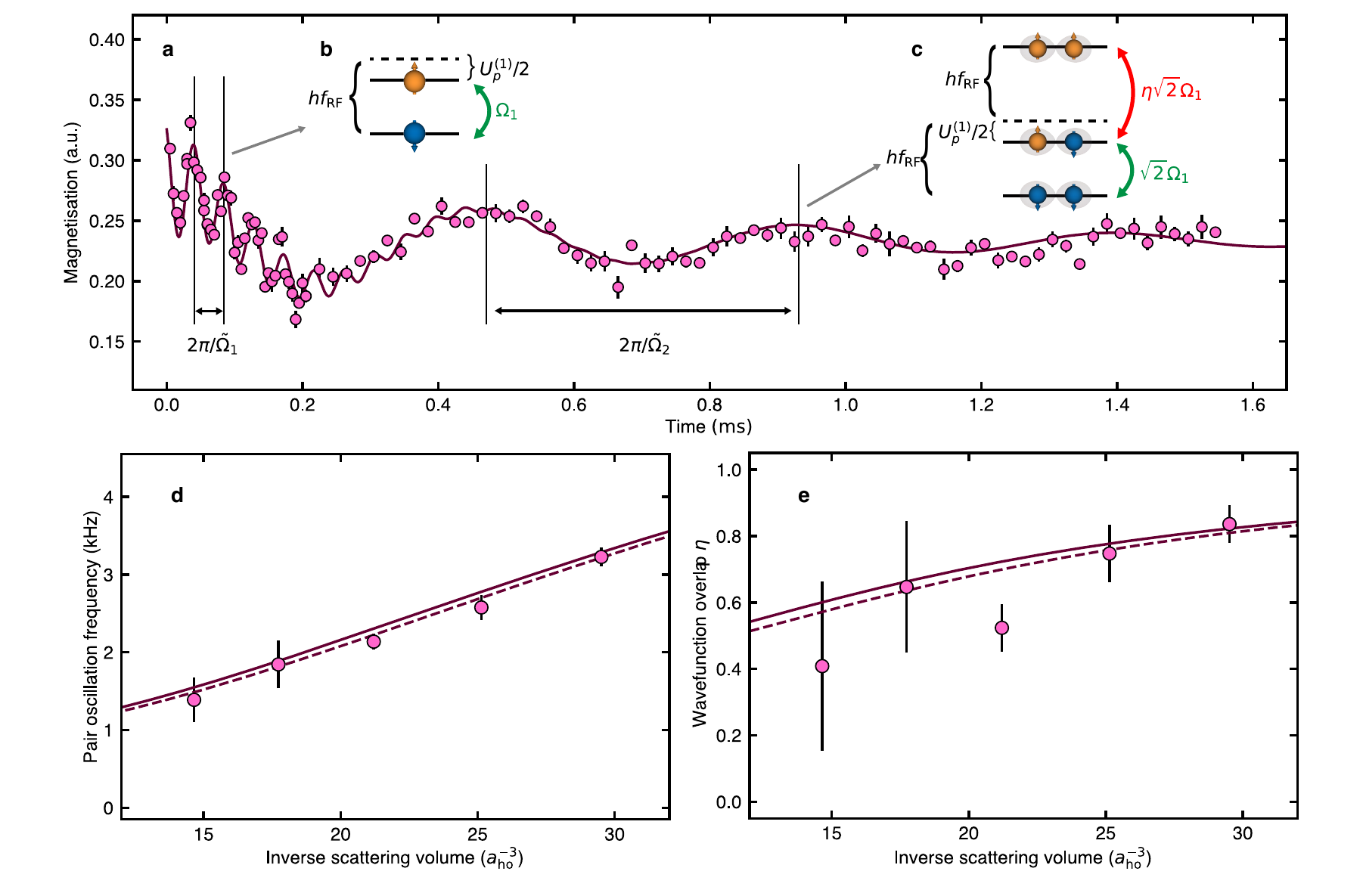}
\caption{{\bf Coherent manipulation of p-wave interacting pairs.} {\bf a}, Temporal oscillations in the magnetisation are observed when applying an RF drive resonant with the two-photon $\ket{\mathrm{T},\downarrow\downarrow}$ to $\ket{\mathrm{T},\uparrow\uparrow}$ transition. Here, the  trap angular frequency is $\omega=2\pi \times 129(2)~\mathrm{kHz}$, and the magnetic field is $200.00(1)~\mathrm{G}$. A boxcar average with a $50~\mu\mathrm{s}$ window is applied to the data for times less than $200~\mu\mathrm{s}$ for visual clarity. Error bars are the standard deviation of repeated data points. A two-frequency fit finds $\widetilde{\Omega}_1=2\pi\times22.7(4)~\mathrm{kHz}$ and $\widetilde{\Omega}_2=2\pi\times2.15(4)~\mathrm{kHz}$, which we identify as one- and two-atom processes, respectively. 
{\bf b}, Rabi oscillations caused by off-resonant coupling of single spins should occur with generalised Rabi frequency $\widetilde{\Omega}_{1}=\{\Omega_1^2+[U_p^{(1)}\!/(2\hbar)]^2\}^{1/2}$. An independent calibration gives the single-particle Rabi frequency $\Omega_1=2\pi\times8.83(2)~\mathrm{kHz}$. 
{\bf c}, Rabi oscillations caused by on-resonant two-photon coupling of $\ket{\mathrm{T},\downarrow\downarrow}$ to $\ket{\mathrm{T},\uparrow\uparrow}$ occur with frequency $\widetilde{\Omega}_2$ (Methods). The coupling strength between $\ket{\mathrm{T},\downarrow\downarrow}$ and $\ket{\mathrm{T},\downarrow\uparrow}$ is $\sqrt{2}\Omega_1$, and the coupling strength between $\ket{\mathrm{T},\downarrow\uparrow}$ and $\ket{\mathrm{T},\uparrow\uparrow}$ is $\eta\sqrt{2}\Omega_1$. 
{\bf d}, The measured pair oscillation frequency $\widetilde{\Omega}_2$ varies with inverse scattering volume. Error bars are the fit uncertainty of the oscillation frequency. The solid and dashed lines are the predictions based on the AbI and PsP calculations respectively (Methods).
{\bf e}, The wavefunction overlap $\eta$ as a function of inverse scattering volume, inferred from the measured two-photon Rabi frequency of {\bf d}. Error bars are the estimated statistical uncertainty of all experimental parameters combined with the fit uncertainty of $\widetilde{\Omega}_2$. The solid and dashed lines are the overlap calculated using AbI and PsP wavefunctions respectively (Methods).}
\label{fig:rabi}
\end{figure*}

Having engineered the required spin symmetry and orbital excitation, we can create and measure strong p-wave interactions via radio-frequency (RF) manipulation. The double-spin-flip resonance condition between $\ket{\mathrm{T},\uparrow\uparrow}$ and $\ket{\mathrm{T},\downarrow\downarrow}$ is $2f_\mathrm{RF}=2f_\mathrm{ZS}+U_p^{(i)}/h$, where $f_\mathrm{RF}$ is the centre frequency of the RF pulse, $f_\mathrm{ZS}$ is the Zeeman splitting of $\uparrow$ and $\downarrow$ spins, and $h$ is Planck's constant. At resonance, the pulse transfers $\ket{\mathrm{T},\downarrow\downarrow}$ pairs to $\ket{\mathrm{T},\uparrow\uparrow}$ through a second-order process via the virtual state $\ket{\mathrm{T},\uparrow\downarrow}$ (see Fig.~\ref{fig:intro}b). Spin flips induced by the RF pulse are detected as changes in the ensemble magnetisation obtained via time-of-flight imaging (Methods). Figure~\ref{fig:intro}c shows repeated measurements with variable $f_\mathrm{RF}$ and features three distinct spin-resonance peaks. The central feature corresponds to flipping an isolated (and thus non-interacting) spin and is used to calibrate the magnetic-field strength. The two side features indicate successful transfers of $\ket{\mathrm{T},\downarrow\downarrow}$ pairs to interacting $\ket{\mathrm{T}, \uparrow \uparrow}$ pair states in $\mathrm{BR}^{(0)}$ and $\mathrm{BR}^{(1)}$ with interaction energies $U_{p}^{(0)}$ and $U_{p}^{(1)}$ respectively. The observed spectra, such as Fig.~\ref{fig:intro}c, constitute the first direct measurements of the elastic p-wave interaction energy. 

\begin{figure*}[!htb]
\centering
\includegraphics[width=1\linewidth]{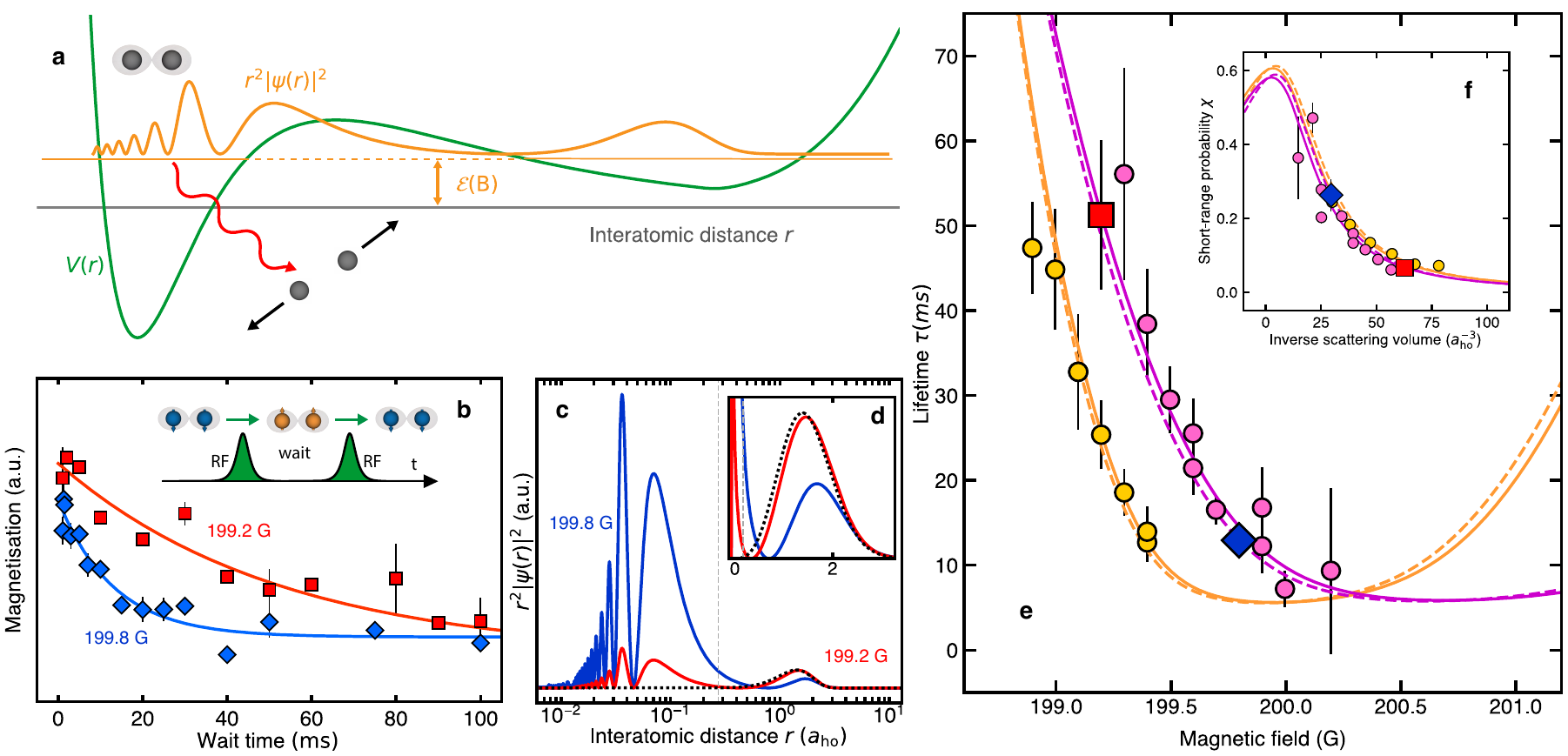}  
\caption{ {\bf Lifetime of p-wave interacting pairs}. 
{\bf a}, The lifetime of an interacting $^{40}$K p-wave pair is limited by dipolar relaxation of the metastable dimer at short range. The Born-Oppenheimer potential energy (green) has a short range interaction component and a long range component due to harmonic confinement (not to scale). 
{\bf b}, The lifetime of $\mathrm{BR}^{(1)}$ pairs is measured as the $1/e$ decay time of the magnetisation for the experimental sequence described in the text (inset). Data is shown for  trap angular frequency $\omega=2\pi\times129(2)\,\mathrm{kHz}$ at magnetic fields $199.20(1)\,\mathrm{G}$ (red) and $199.80(1)\,\mathrm{G}$ (blue) with lifetimes 51(9)\,ms and 13(2)\,ms respectively. Error bars show the standard deviation of repeated measurements. 
{\bf c}, The probability density $r^2 |\psi|^2$ exhibits distinct short- and long-range components. Coloured solid lines correspond to ab-initio wavefunctions for experimental conditions in panel {\bf b}; the vertical dashed line indicates the distance up to which the short-range probability $\chi$ is calculated (see Supplements). 
{\bf d}, Probability densities of {\bf c} with linear scaling in interatomic distance. The black dotted wavefunctions give the non-interacting $n_{\mathrm{rel}}=1$ oscillator state for comparison.  
{\bf e}, The measured lifetime decreases for magnetic fields closer to the unitary limit. Shown are data for harmonic  trap angular frequencies $\omega=2\pi\times90(2)\,\mathrm{kHz}$ (yellow) and $\omega=2\pi\times129(2)\,\mathrm{kHz}$ (pink). The square and diamond markers correspond to the temporal data shown in {\bf b}. Error bars are the fit uncertainty of the $1/e$ decay time. The lines are theory predictions based on the AbI wavefunctions (solid) and PsP wavefunctions (dashed). {\bf f}, The variation of the measured short-range probability $\chi$ with inverse scattering volume. Solid and dashed lines are predictions from AbI and PsP calculations respectively where $\chi$ attains a maximum value at the unitary point (Methods). 
}
\label{fig:lifetime}
\end{figure*}

We probe the eigenspectrum of interacting p-wave atoms through RF spectroscopic scans at variable trap frequency and magnetic fields. The measured energies test the validity of an analytical treatment that uses the p-wave pseudopotential  \cite{Blume04,Idziaszek} (PsP) to calculate the interaction energy as a function of the energy-dependent scattering volume, $v(\mathcal{E})$ (Methods). At unitarity, $v(\mathcal{E})$ diverges but the interaction energy remains finite with $U_p^{(0)} = -\hbar \omega$ and $U_p^{(1)} = +\hbar \omega$. This resonant behaviour is universal, independent of any microscopic details of the atomic collisions. In Figs.~\ref{fig:energy}a-c we compare the measured interaction energy to the PsP prediction, including a leading-order anharmonic correction (Methods). In both branches, we observe agreement across a wide range of interaction strengths -- including in the unitary limit. Figure~\ref{fig:energy}d collects all data as $\mathcal{E}/\hbar\omega$ versus $v(\mathcal{E})/a_\mathrm{ho}^3$, where $a_{\mathrm{ho}} = \sqrt{{\hbar}/{\mu \omega}}$ is the harmonic oscillator  length  and $\mu=m/2$ is the reduced mass. The data collapse demonstrates the exclusive dependence of p-wave interaction energies on a single parameter, which implies the the universal applicability of this result to any p-wave interacting system in the tight-binding limit.

Further insight is provided by comparing the wavefunctions of the PsP theory to those obtained numerically from an {\em ab-initio} (AbI) interaction potential specific to $^{40}$K (see Figs.~\ref{fig:energy}e,f,g). At short length scales, $r \lesssim 0.1 a_\mathrm{ho}$, the PsP diverges while the AbI does not. However as described in Supplements, after regularisation with a short-range cutoff (at the van der Waals length), the PsP wavefunction is normalisable and accurately predicts the long-range wavefunction. 
Far from resonance, both the PsP and AbI wavefunctions match the non-interacting oscillator states ($n_{\mathrm{rel}}=1$ in Fig.~\ref{fig:energy}g and $n_{\mathrm{rel}}=3$ in ~\ref{fig:energy}e). At unitarity (Fig.~\ref{fig:energy}f), the wavefunction asymptotically resembles that of a non-interacting bosonic pair in the $n_\mathrm{rel}=2$ motional state, since an exchange-odd wavefunction with an additional $\pi/2$ scattering phase imitates a non-interacting exchange-even wavefunction. 

Next, we demonstrate coherent manipulation of p-wave interacting pairs, which also probes the interacting wavefunctions. As shown in Fig.~\ref{fig:rabi}a, application of RF radiation under the two-photon resonance condition for the $\mathrm{BR}^{(1)}$ branch results in an oscillating ensemble magnetisation with a two-tone frequency character. The faster oscillation evident at short times corresponds to (off-resonant) $\uparrow$-to-$\downarrow$ Rabi oscillations of single spins; the slower oscillation persisting for longer time corresponds to resonant Rabi oscillations of pairs between  $\ket{\mathrm{T},\downarrow\downarrow}$ and $\ket{\mathrm{T},\uparrow\uparrow}$. 
The oscillation frequency of the pairs is sensitive to the wavefunction overlap $\eta$ between the interacting and non-interacting states (Methods). The two-atom RF Rabi frequency also has a $\sqrt{2}$ enhancement above the single-atom coupling $\Omega_1$ due to constructive interference among pure spin-triplet states. 
In Fig.~\ref{fig:rabi}d, we compare the observed pair Rabi frequency to theoretical predictions for a range of inverse scattering volumes, and find excellent agreement. This measurement allows us to directly extract $\eta$, as shown in Fig.~\ref{fig:rabi}e. The observed agreement between theory and experiment demonstrates coherent control of the system and success of both the AbI and regularised PsP to predict the interacting wavefunction.

A final experiment probes the lifetime of the p-wave interacting pairs. In the absence of three-body recombination, the lifetimes are limited for $^{40}$K by inelastic two-body collisions of pairs of atoms at short interatomic separation (see Fig.~\ref{fig:lifetime}a), with a characteristic lifetime $\tau_d \approx 3.4$\,ms \cite{Ticknor:2004,chipresonance} (Methods). The pair lifetime is measured with a double-pulse sequence (see inset of Fig.~\ref{fig:lifetime}b) in which  $\ket{\mathrm{T},\uparrow\uparrow}$ pairs are created, held for a variable hold time, and transferred back to $\ket{\mathrm{T},\downarrow\downarrow}$. The survival lifetime is extracted from the exponential decay of the ensemble magnetisation, as shown in Fig.~\ref{fig:lifetime}b for two different magnetic fields. 
Even though the strong lattice confinement has increased atomic density, we find lifetimes in excess of 50\,ms, which is fifty times longer than previously observed for free-space dimers \cite{Gaebler:2007}. The relatively long lifetime of the 199.2\,G condition can be understood by its reduced  probability ($\chi$) for small inter-nuclear separation $r$, where relaxation processes are strongest (see Figs.~\ref{fig:lifetime}c,d). Both AbI and regularized PsP wavefunctions allow us to calculate $\chi$; as shown in Fig.~\ref{fig:lifetime}e,  these show excellent agreement with measured lifetimes using the simple relation $\tau = \tau_d/\chi$. The observed agreement across all interaction energies, demonstrates the full suppression of three-body recombination, the absence of band relaxation, the validity of both the ab-initio and PsP wavefunctions, and the calculation of $\tau_d$. Figure~\ref{fig:lifetime}f plots $\chi$ versus $a_\mathrm{ho}^3/v(\mathcal{E})$, which emphasises the applicability of the wavefunctions to any p-wave system, even those (such as $^6$Li \cite{Zhang:2004cy,chevy2005,Schunck:2005cf,Fuchs:2008ka,Inada:2008hz,Nakasuji:2013gw,Waseem:2016ki,Waseem:2017ep}) without the dipolar relaxation channel present in $^{40}$K $\ket{\mathrm{T}, \uparrow\uparrow}$ pairs \cite{Ticknor:2004}. 

The observation, control, and comprehensive understanding of strong p-wave interactions demonstrated here illuminates a path towards the assembly of new many-body states of matter. In a full lattice model, the measured $U_p^{(i)}$ calibrates the on-site interaction, while lattice depth controls tunnelling between sites. For sufficiently small tunnelling strength, losses might continue to be suppressed either through the quantum Zeno effect \cite{han2009PRL} or by the energetic gaps to triple on-site occupation. In two dimensions, the $U^{(0)}_p<0$ interactions observed here in $\mathrm{BR}^{(0)}$ in the $|M_L| = 1$ channel are the pre-cursors of $p+ip$ superfluidity \cite{Gurarie2007,Volovik,Mizushima2016,Botelho2005}, which features non-trivial topological properties, as well as gapless chiral edge modes, or ``Majorana zero modes'' in vortex cores \cite{Nayak2008,Alicea2012}. These are non-abelian anyons that are predicted to offer unique opportunities for topological quantum computation and robust quantum memories \cite{Alicea2012,Nayak2008,Ivanov2001}. Even a metastable many-body state would allow for the study of topological states in a quenched p-wave superfluid \cite{foster2014PRL}. The $U_p^{(1)} > 0$ interactions observed here are the pre-cursors of orbital magnetism known from transition metal oxides \cite{tokura2000orbital}, as well as orbitally ordered Mott insulators \cite{imada1998insulator,khaliullin2005insulator} in a multi-band Fermi-Hubbard model \cite{mamaevpwave}. Strong orbital interactions demonstrated in this work can also be used to engineer low-entropy states in a multi-band lattice system \cite{BakrOEB} and a full gate-based control of entangled many-body states \cite{Mamaev2020}. Finally, the universal nature of the observed interaction energies indicates that it would be reproduced in other ultracold p-wave systems such as $^6$Li \cite{Zhang:2004cy,Schunck:2005cf,Fuchs:2008ka,Inada:2008hz,Nakasuji:2013gw,Waseem:2016ki,Waseem:2017ep} and ultracold fermionic molecules \cite{Ospeklaus2010,Petrov2018,Duda2022}.

\begin{acknowledgments}
We acknowledge insightful discussions with Fr\'ed\'eric Chevy and Shizhong Zhang, and helpful manuscript comments from John Bohn, Adam Kaufman, and Robyn Learn. This work is supported by the AFOSR grants FA9550-19-1-0275, FA9550-19-1-7044, and FA9550-19-1-0365, by ARO W911NF-15-1-0603, by the NSF's JILA-PFC PHY-1734006 and PHY-2012125 grants, by NIST, and by NSERC. 
\end{acknowledgments}

\bibliography{pwavebiblio.bib}

\clearpage
\newpage

\appendix
\section{Methods \label{sec:Methods}}

%%%%%
\noindent{\bf Spin and motional wavefunctions.} 
%%%%%
The single-atom spin states $\uparrow$ and $\downarrow$ used in the experiment are the $m_f = -9/2$ and $m_f = -7/2$ spin states of the ground hyperfine manifold of $^{40}$K with total spin $f=9/2$. The pair spin wavefunctions are given by
$\ket{\mathrm{S},\uparrow\downarrow}=\left(\ket{\downarrow,\uparrow}-\ket{\uparrow,\downarrow}\right)/\sqrt{2}$, $\ket{\mathrm{T},\uparrow\uparrow}=\ket{\uparrow,\uparrow}$, $\ket{\mathrm{T},\downarrow\downarrow}=\ket{\downarrow,\downarrow}$, and $\ket{\mathrm{T},\uparrow\downarrow}=\left(\ket{\downarrow,\uparrow}+\ket{\uparrow,\downarrow}\right)/\sqrt{2}$.
The motional states of the pair are defined in terms of spherical harmonic oscillator eigenstates for the relative atomic separation $r$ (see Supplements),
\begin{equation}
\begin{aligned}
    \ket{E} &= \ket{\mathcal{N},L,M_L},\\
    E &= \hbar \omega\left(2\mathcal{N} + L + \frac{3}{2}\right).
\end{aligned}
\end{equation}
Here $\mathcal{N} \in \{0,1,2,\dots\}$ is the radial excitation number, $L \in \{0,1,2,\dots\}$ is the relative angular momentum, and $M_L \in \{-L,\dots, L\}$ the angular momentum projection along the magnetic field axis. The total number of motional excitations can also be characterised by a single quantum number $n_{\mathrm{rel}}=2\mathcal{N}+L=1,3,\dots$, since $L=1$ for $p$-wave interactions. 

%%%%%
\bigskip \noindent{\bf State preparation and readout.} 
%%%%%
The degenerate Fermi gas is a balanced spin mixture of $^{40}$K in its lowest two hyperfine spin states 
% ($\downarrow$ and $\uparrow$) 
created via sympathetic optical evaporation with $^{87}$Rb in a crossed optical dipole trap \cite{edgelattice,rhyslattice}. After evaporation, the gas typically contains $2 \times 10^5$ atoms with temperature $0.1T_F$, where $T_F$ is the Fermi temperature. 

The optical lattice potential is formed by orthogonal retro-reflected laser beams of wavelengths $\lambda_{xy}=1054~\mathrm{nm}$ in the $xy$ plane and $\lambda_z=1064~\mathrm{nm}$ along the $z$-axis with beam waists $\left(w_x,w_y,w_z\right)=\left(60, 60, 85\right)$\,$\mu$m.  The potential depth of the lattice is parameterised in terms of the  recoil energy of the $xy$ lattice beams $E_{\mathrm{R}}=\hbar^2 k_L^2/2m$, where $k_L=2\pi/\lambda_{xy}$ and $m$ is the mass of a $^{40}$K atom.  The harmonic trap angular frequency of a lattice site $\omega$ is given by $\hbar \omega = E_{\mathrm{R}}\sqrt{4V_\mathrm{L}/E_{\mathrm{R}}}$ where $V_\mathrm{L}$ is the lattice depth. The lattice depths are regulated to be isotropic and are verified by comparing amplitude-modulation spectroscopy to band structure. We estimate the lattice anisotropy to be less than $2\%$. 

Isolated pairs of atoms in the $\ket{\mathrm{S}, \uparrow \downarrow} \ket{0}_{\text{rel}}$ state are created by  ramping the lattice depth to 10$E_{\mathrm{R}}$ in 150\,ms, waiting for 50\,ms, and then  suppressing tunnelling with a fast ramp to 60$E_{\mathrm{R}}$ in 250\,$\mu\mathrm{s}$.  In-situ fluorescence imaging with a quantum gas microscope verifies that approximately 10\% of the sites are doubly occupied. The lattice depth is then ramped to 200$E_{\mathrm{R}}$ in 100\,ms, and the magnetic field along the $z$ lattice direction is ramped to 197\,G in 150\,ms. Atom pairs in the $\ket{\mathrm{S}, \uparrow \downarrow}\ket{0}_{\mathrm{rel}}$ state are transferred to the $\ket{\mathrm{T}, \downarrow \downarrow} \ket{1}_{\text{rel}}$ state by a $65~\mu\mathrm{s}$ Raman $\pi$-pulse which is detuned from the Zeeman splitting by a motional quanta and the on-site s-wave interaction energy of the $\ket{\mathrm{S}, \uparrow \downarrow}\ket{0}_{\mathrm{rel}}$ state.

To perform state readout, the magnetic field is first ramped (in 50\,ms) to 195\,G  where the atom pairs are weakly interacting.  The resultant absolute spin populations of the $\uparrow$ and $\downarrow$ states are measured via absorption imaging after band mapping and a 15\,ms time of flight. A double shutter imaging technique enables  measurement of both spin populations in a single experimental realisation.

%%%%%
\bigskip \noindent{\bf Raman excitation.} 
%%%%%
The Raman coupling is generated by two linearly polarised beams in the $xy$ plane whose propagation directions are oriented at $30^{\circ}$ and $60^{\circ}$ respectively with the $x$ and $y$ lattice directions. A small angular deviation from the $xy$ plane allows excitations along the $z$ motional degree of freedom, and thus $M_L = 0$ features are present in the spectra. The single-photon detuning of each Raman laser beam is stabilised to $-50.1\,\mathrm{GHz}$ from the D2 transition and is chosen to avoid undesired photo-association of pairs of $^{40}$K atoms  \cite{39KPA} at a single site.

%%%%%
\bigskip \noindent{\bf RF spectroscopy.} 
%%%%%
After preparing the non-interacting $\ket{\mathrm{T}, \downarrow \downarrow}\ket{1}_{\mathrm{rel}}$ pair state, the lattice depth and magnetic field are ramped sequentially in 50\,ms to their operating values as indicated in the main text. The radio-frequency spectroscopy implements the hyperbolic secant (HS1) pulse shape which is defined by the following time-dependent detuning $\delta(t)$ about the central frequency $f_{\mathrm{RF}}$, and Rabi frequency $\Omega(t)$:
\begin{align}
    \Omega(t) &= \Omega_0\sech\left(2\beta\frac{t}{T_\mathrm{p}}\right)\\
    \delta(t) &=\delta_\mathrm{c}+\delta_\mathrm{m}\tanh\left(2\beta\frac{t}{T_\mathrm{p}}\right)\,.
\end{align}
Here, $\Omega_0$ is the peak Rabi frequency at resonance, which is essentially the single-particle Rabi frequency $\Omega_1$. Note that in the Rabi-oscillation measurements, the Rabi frequency is fixed as a constant of $\Omega(t) = \Omega_1$. In the expression of the detuning above, $\delta_\mathrm{m}$ is the maximum absolute detuning with respect to the central detuning of $\delta_\mathrm{c}/(2 \pi) = f_{\mathrm{RF}}-f_{\mathrm{ZS}}$, and $T_\mathrm{p}$ is the characteristic pulse time. The dimensionless tuning parameter $\beta$ sets the relative sharpness of the sweep. Typical experimental parameters are $\delta_\mathrm{m}=2\pi \times 2.5\,\mathrm{kHz}$, $\Omega_0 = 2\pi \times 8.8\,\mathrm{kHz}$, $\beta=0.05$, and $T_\mathrm{p}=2\,\mathrm{ms}$.

%%%%%
\bigskip \noindent{\bf Feshbach resonance.} 
%%%%%
In free space, $\ket{\mathrm{T},\uparrow\uparrow}$ pairs of atoms have a p-wave magnetic Feshbach resonance at 198.30\,G for $M_L=\pm 1$, and 198.80\,G for $M_L=0$. In the effective range approximation, the energy dependent scattering volume $v(\mathcal{E})$ for each collisional channel is given by 
\begin{equation}
    v(\mathcal{E}) \approx \left[\frac{1}{v_\mathrm{bg}\left(1-\frac{\Delta}{B-B_0}\right)} + \frac{\mu \mathcal{E}}{\hbar^2 R(B)}\right]^{-1},
\end{equation}
where $v_{\mathrm{bg}}$ is the background scattering volume, $\Delta$ is the resonance width, $B_0$ is the resonant magnetic field, $B$ is the applied magnetic field, $\mu=m/2$ is the reduced mass, and $R(B)$ is the field dependent effective range given by the linear expression $R(B)=R_0[1+ (B-B_0)/\Delta_r]$. The resonance parameters for $M_L=0$ are $v_{\mathrm{bg}}=-(108.0a_0)^3$, $\Delta=-19.89~\mathrm{G}$, $R_0=49.4a_0$, and $\Delta_r=21.1~\mathrm{G}$. The resonance parameters for $M_L=\pm1$ are $v_{\mathrm{bg}}=-(107.35a_0)^3$, $\Delta=-19.54~\mathrm{G}$, $R_0=48.9a_0$, and $\Delta_r=21.7~\mathrm{G}$ \cite{chipresonance}.

%%%%%
\bigskip \noindent{\bf Pseudopotential.} 
%%%%%
The p-wave interaction between two identical atoms can be computed via a regularised pseudopotential  \cite{Blume04,Omant77,Idziaszek} given by 
\begin{equation} V_p(\mathbf{r}) = \frac{12 \pi \hbar^2 v(\mathcal{E})}{m} \overleftarrow{\nabla}_{\mathbf{r}} \delta^{(3)}(\mathbf{r})\overrightarrow{\nabla}_{\mathbf{r}} \frac{1}{2}\frac{\partial^2}{\partial r^2} r^2,
\label{eq:pseudo}
\end{equation}
where  $\mathbf{r}$ is the relative position of the atoms and $r=|\mathbf{r}|$ their separation, $\delta^{(3)}$ is the 3D Dirac delta function, and $\overleftarrow{\nabla}_{\mathbf{r}}$, $\overrightarrow{\nabla}_{\mathbf{r}}$ are left-, right-acting gradients respectively. %
In principle, the energy-dependent scattering volume $v(\mathcal{E})$ is different for the $M_{L}=\pm 1$ and $M_L = 0$ channels due to dipolar interactions. Thus, the pseudopotential should be separated into terms with spatial derivatives acting in the $x-y$ plane and $z$ direction (since the magnetic field points along $z$) with different scattering volumes. However, this does not lead to coupling between the $M_L$ channels. Therefore, the energies of the different channels are simply given by the solution of the isotropic case with the appropriate scattering volume.

An isotropic scattering volume permits an analytic solution for the energy $\mathcal{E}$, which is given implicitly by \cite{Blume04,Idziaszek}
\begin{equation}
\label{eq_EnergyImplicitEquation}
    -\frac{a_{\mathrm{ho}}^3}{v(\mathcal{E})} = 8\frac{\Gamma\left(\frac{5}{4} - \frac{\mathcal{E}}{2\hbar\omega}\right)}{\Gamma\left(-\frac{1}{4} -\frac{ \mathcal{E}}{2\hbar\omega}\right)}
\end{equation}
where $\Gamma(z)$ is the Gamma function. The corresponding spatial wavefunction can be written as \cite{Blume04}
\begin{equation}
    \psi_{\mathrm{int}}(r) = \begin{cases} 
      \mathcal{A}\> \frac{r}{a_{\mathrm{ho}}}\> e^{-\frac{r^2}{2 a_{\mathrm{ho}}^2}}\> U\left(-\frac{\mathcal{E}}{2\hbar \omega}+ \frac{5}{4},\frac{5}{2},\frac{r^2}{a_{\mathrm{ho}}^2}\right) & r > r_{\mathrm{cut}} \\
      0 & r \leq r_{\mathrm{cut}}
   \end{cases}
\end{equation}
where $\mathcal{A}$ is a normalisation constant, $U(a,b,z)$ is the confluent hypergeometric function of the second kind, and $r_{\mathrm{cut}} = 50 a_0$ is a cutoff used to treat the divergence as $r \to 0$, obtained by comparing directly to ab-initio wavefunction calculations (see Supplements).

%%%%%
\bigskip \noindent{\bf Anharmonic corrections.} 
%%%%%
The anharmonic correction to the pseudopotential energy is approximated using first-order perturbation theory. We compute the expectation value of fourth-order Taylor expansion terms of the lattice trapping potential about the center of a lattice site (see Supplements). The resulting correction is
\begin{widetext}
\begin{equation}
\Delta \mathcal{E}_{\mathrm{anharmonic}} =-E_R \left(\frac{1}{10a_{\mathrm{ho}}^4}\int_{0}^{\infty}dr \text{ } r^6 |\psi_{\mathrm{int}}(r)|^2 +\frac{1}{2a_{\mathrm{ho}}^2} \int_{0}^{\infty}dr \text{ } r^4 |\psi_{\mathrm{int}}(r)|^2 - \frac{17}{8} \right).
\end{equation}
\end{widetext}

%%%%%
\bigskip \noindent{\bf Pair Rabi oscillations.} 
%%%%%
The Rabi oscillation spin dynamics of an interacting pair is captured by the following three-level model (see Supplementary),
\begin{equation}
    \hat{H}^{\mathrm{pair}} = \left(\begin{array}{ccc} 0 & \frac{\hbar \Omega_1}{\sqrt{2}} & 0 \\ \frac{\hbar\Omega_1}{\sqrt{2}} & -\frac{U_p^{(1)}}{2} & \frac{\hbar\Omega_1}{\sqrt{2}}\eta \\
    0 & \frac{\hbar\Omega_1}{\sqrt{2}}\eta & 0\end{array}\right),
\end{equation}
written in the basis of $\{\ket{T,\downarrow\downarrow}\ket{1}_{\mathrm{rel}}$,  $\ket{T,\uparrow\downarrow}\ket{1}_{\mathrm{rel}}$,  $\ket{T,\uparrow\uparrow}\ket{\psi_{\mathrm{int}}}_{\mathrm{rel}}\}$. Here $\Omega_1$ is the single-photon Rabi frequency of the RF drive, while $\eta$ is a spatial wavefunction overlap between the non-interacting and interacting states (see Supplements),
\begin{equation}
\begin{aligned}
    &\eta = {}_{\mathrm{rel}}\!\bra{\psi_{\mathrm{int}}} 1\rangle_{\mathrm{rel}} =  \int_{0}^{\infty}dr \> r^2 \psi_{\mathrm{int}}^{*}(r)\psi_{\mathrm{rel}}^{(n_{\mathrm{rel}}=1)}(r), \\
    &\mbox{where} \\
   &\psi_{\mathrm{rel}}^{(n_{\mathrm{rel}}=1)}(r)=\left({\frac{8}{3 \pi^{1/2} a_{\mathrm{ho}}^3}}\right)^{1/2} \frac{r}{a_{\mathrm{ho}}} \exp{\left(-\frac{r^2}{2 a_{\mathrm{ho}}^2}\right)}\,.
\end{aligned}
\end{equation}
In the limit of $U_{p}^{(1)} \gg \hbar \Omega_1$, dynamics under this Hamiltonian is characterised by a single frequency
\begin{equation}
  \widetilde{\Omega}_2 = \frac{\sqrt{\left(U_p^{(1)}\!/\hbar\right)^2 + 8(1+\eta^2) \Omega_1^2}- U_p^{(1)}\!/\hbar}{4}.
\end{equation}
Experimental measurements extract $\eta$ from the above equation, since all other parameters are independently measured.

%%%%%
\bigskip \noindent{\bf Lifetime prediction.} 
%%%%%
The lifetime $\tau$ of the interacting state is limited by inelastic decay due to dissociation of the pair into unbound atoms. Dipolar interactions couple the interacting state $\ket{T,\uparrow\uparrow}\ket{1}_{\mathrm{rel}}\ket{0}_{\mathrm{CM}}$ to a lossy dimer state at short interatomic separation, which undergoes dissociation with a characteristic lifetime $\tau_d$. The dimer lifetime for $M_L = +1$ and $M_L = -1$ is $\tau_{+1}=8.7$ ms and $\tau_{-1}=2.1$ ms respectively \cite{chipresonance}. Our motional excitation is predominantly along a single Cartesian lattice direction in the $\hat{x}$-$\hat{y}$ plane, which corresponds to an equal superposition of $M_{L} = +1, -1$; the characteristic lifetime is thus $\tau_d^{-1} = (\tau_{+1}^{-1} + \tau_{-1}^{-1})/2$, such that $\tau_d = 3.4$\,ms. The actual lifetime further depends on the short-range wavefunction probability $\chi$. We theoretically predict $\chi$ from the interacting wavefunctions by computing the overall probability up to a characteristic threshold (see Supplements). At all probed magnetic fields, we see a clear distinction between short- and long-range components, such as in Fig.~\ref{fig:lifetime}c. The threshold is chosen to capture the short-range portion of the wavefunction only.

\section{Supplemental Information}

%%%%%%%%%%%%%%
\bigskip \noindent{\bf Interacting wavefunction calculations.}
%%%%%%%%%%%%%%
The spatial wavefunction $\ket{\mathcal{N},L,M_L}$ for a non-interacting spherical harmonic oscillator state is
\begin{equation}
\begin{aligned}
     \psi_{\mathcal{N},L,M_L}(r) =& \mathcal{A}_0  \left(\frac{r}{a_{\mathrm{ho}}}\right)^{L}e^{-\frac{r^2}{2 a_{\mathrm{ho}}^2}} \mathcal{L}_{\mathcal{N}}^{(L+1/2)}\left(\frac{r^2}{a_{\mathrm{ho}}^2}\right) \\ &\times\;Y_{L,M_L}(\theta_r,\phi_r),
     \end{aligned}
\end{equation}
where $\mathcal{A}_0$ is a normalisation constant, $\mathcal{L}_{\mathcal{N}}^{(L+1/2)}$ is the generalised Laguerre polynomial, and $Y_{L,M_L}$ is the spherical harmonic function. Here $(r,\theta_r,\phi_r) \equiv \vec{r}$ is the relative separation of the atoms in spherical coordinates, with $\vec{r}=\vec{r}_1-\vec{r}_2$ for individual atom positions $\vec{r}_1$, $\vec{r}_2$. There is a corresponding center-of-mass position $(R,\theta_R,\phi_R) \equiv \vec{R}=(\vec{r}_1 +\vec{r}_2)/2$, although this is unaffected by interactions and decouples.

The total number of motional excitations can also be characterised by a single quantum number $n_{\mathrm{rel}}=2\mathcal{N}+L$ which we use in the main text. There is a corresponding center-of-mass overall excitation number $n_{\mathrm{CM}}$, which we assume to be constant and set to $n_{\mathrm{CM}}=0$ throughout. Since we study identical fermions, the relative wavefunction has odd parity under particle exchange. This prevents $s$-wave ($L=0$) collisions and fixes $L$ to be odd to have overall exchange antisymmetry (since odd-$L$ spatial wavefunctions have odd parity). This restriction in turn mandates that the overall motional excitation number is odd, $n_{\mathrm{rel}}=1,3,\dots$. We must thus inject a motional quanta into the state prepared in the ground state of the system to study the $p$-wave interactions. This is done by our Raman pulse, which creates a single excitation along a Cartesian direction of the lattice. The angular momentum is $L=1$ ($p$-wave interactions) and the projection can be $M_L =\pm 1$ (for an excitation along $\hat{x}-\hat{y}$) or $M_L = 0$ (for an excitation along $\hat{z}$). The non-interacting oscillator energy of the relative coordinate is initially $E=\frac{5}{2}\hbar \omega$; all interaction energy shifts are measured relative to this initial value.

Turning to the interacting pair, the wavefunction for two identical fermionic atoms in a harmonic trap interacting via the $p$-wave pseudopotential can be written analytically as \cite{Blume04}
\begin{equation}
    \psi_{\mathrm{pseudo}}(r) = \mathcal{A} \frac{r}{a_{\mathrm{ho}}} e^{-\frac{r^2}{2 a_{\mathrm{ho}}^2}}\> U\left(-\frac{\mathcal{E}}{2\hbar \omega}+ \frac{5}{4},\frac{5}{2},\frac{r^2}{a_{\mathrm{ho}}^2}\right),
\end{equation}
where $\mathcal{E}$ is the energy of the system (including the interaction energy and harmonic trap energy of the relative coordinate), while $U(a,b,z)$ is the confluent hypergeometric function of the second kind. Unlike the $s$-wave case \cite{busch1998sWave}, this wavefunction is not normalisable due to its $\sim 1/r^2$ divergence as $r \to 0$. However, we find that the wavefunction can still accurately reproduce the long-range physics of the problem when a short-range cutoff $r_{\mathrm{cut}}$ is imposed. Specifically, we define
\begin{equation}
\label{eq_ExactWavefunctionPseudopotential}
    \psi_{\mathrm{int}}(r) = \begin{cases} 
      \psi_{\mathrm{pseudo}}(r) & \mbox{for}\quad r > r_{\mathrm{cut}} \\
      0 & \mbox{for}\quad r \leq r_{\mathrm{cut}}
   \end{cases}
\end{equation}
With this cutoff, the normalisation constant can be numerically computed via
\begin{equation}
    \mathcal{A}^{-2} = \int_{r_{\mathrm{cut}}}^{\infty}\!\!dr \, r^2  \left[ \frac{r}{a_{\mathrm{ho}}}\> e^{-\frac{r^2}{2a_{\mathrm{ho}}^2}}\>U\left(-\frac{\mathcal{E}}{2\hbar \omega}+ \frac{5}{4},\frac{5}{2},\frac{r^2}{a_{\mathrm{ho}}^2}\right)\right]^2.
\end{equation}

The overall amplitude of the wavefunction (and thus the long-range properties) depends strongly on the chosen cutoff $r_{\mathrm{cut}}$. To determine the correct value of $r_{\mathrm{cut}}$ that captures the physics seen in the experiment and test the validity of the pseudopotential, we numerically compute wavefunctions for the same problem using ab-initio molecular calculations.

\begin{table}[tb]
\centering
\begin{tabular}{ c c }
 $\text{Channel }\sigma$ & \text{Spin basis state }$\ket{\phi_{\sigma}}$\\ \hline
 $bb$ & $\Ket{\frac{9}{2},-\frac{7}{2}}_{1} \Ket{\frac{9}{2},-\frac{7}{2}}_2$ \\  
 $ac$ & $\frac{1}{\sqrt{2}}\left(\Ket{\frac{9}{2},-\frac{5}{2}}_{1}\Ket{\frac{9}{2},-\frac{9}{2}}_{2}+\Ket{\frac{9}{2},-\frac{9}{2}}_{1} \Ket{\frac{9}{2},-\frac{5}{2}}_{2}\right)$ \\
 $aq$ & $\frac{1}{\sqrt{2}}\left(\Ket{\frac{9}{2},-\frac{9}{2}}_{1}\Ket{\frac{7}{2},-\frac{5}{2}}_{2}+\Ket{\frac{7}{2},-\frac{5}{2}}_{1}\Ket{\frac{9}{2},-\frac{9}{2}}_{2}\right)$ \\
 $br$ & $\frac{1}{\sqrt{2}}\left(\Ket{\frac{9}{2},-\frac{7}{2}}_{1}\Ket{\frac{7}{2},-\frac{7}{2}}_{2}+\Ket{\frac{7}{2},-\frac{7}{2}}_{1}\Ket{\frac{9}{2},-\frac{7}{2}}_{2}\right)$ \\
 $rr$ & $\frac{1}{\sqrt{2}}\left(\Ket{\frac{7}{2},-\frac{7}{2}}_{1}\Ket{\frac{7}{2},-\frac{7}{2}}_{2}+\Ket{\frac{7}{2},-\frac{7}{2}}_{1}\Ket{\frac{7}{2},-\frac{7}{2}}_{2}\right)$ \\
\end{tabular}
\caption{Collisional channels $\sigma$ and corresponding symmetric spin basis states $\ket{\phi_{\sigma}}$. Here $\ket{f,m_f}_{\alpha}$ is the state of atom $\alpha \in 1,2$ with total spin $f$ and spin projection $m_f$.}
\label{tab_CollisionChannels}
\end{table}

Our ab-initio calculations \cite{chapurin2019PRL}, which neglect the magnetic-dipole interaction, assume that the two colliding $^{40}$K atoms have total spin projection $M_F = -7$. The total wavefunction is expressed in a basis of symmetric hyperfine spin states $\{\ket{\phi_{\sigma}}\}$ corresponding to different collisional channels $\sigma$. Table~\ref{tab_CollisionChannels} shows the channel labels and corresponding spin states, with $\ket{f,m_f}_{\alpha}$ being the spin state of atom $\alpha \in 1,2$ with atomic hyperfine spin $f$ and projection $m_f$. Our calculations consist of a fully coupled channel approach \cite{chapurin2019PRL} 
where realistic singlet and triplet Born-Oppenheimer potentials \cite{falke2008PRA} are used along with a harmonic trap. The full problem is diagonalised to find the lowest eigenvalues and eigenstates for trap depths and magnetic field values as used in the experiment. Since our model neglects magnetic dipole interactions, we cannot directly distinguish between the $|M_L|=1$ and $M_L = 0$ components, whose free-space Feshbach resonance positions differ by $\approx 0.5$ G. For that reason, we apply an overall magnetic field shift to align the lowest resulting branch $\text{BR}^{(0)}$ with the corresponding branch of the pseudopotential.

For each branch $\text{BR}^{(i)}$, the output of these calculations is a wavefunction of the form
\begin{equation}
    \ket{\psi_{\mathrm{ab-initio}}} = \sum_{\sigma} \psi_{\sigma}(r) \ket{\phi_{\sigma}},
\end{equation}
where $\psi_{\sigma}(r)$ is the spatial wavefunction for spin channel $\sigma$. The wavefunction is normalised via
\begin{equation}
    \sum_{\sigma} \int_0^{\infty}dr\>r^2[\psi_{\sigma}(r)]^2 = 1.
\end{equation}
Only the $\psi_{bb}(r)$ component has non-negligible amplitude at long range $r \gtrsim a_{\mathrm{ho}}$. This component adiabatically maps to the $n_{\mathrm{rel}}=1,3,\dots$ non-interacting oscillator states that the system resides in when far from Feshbach resonance, and corresponds to both atoms being in the $\uparrow$ spin state. The pseudopotential wavefunction should thus match the $bb$ component in the long range regime,
\begin{equation}
    \psi_{\mathrm{int}}(r) \approx \psi_{bb}(r) \text{ for $r \gtrsim a_{\mathrm{ho}}$.}
\end{equation}
By comparing the wavefunctions for different values of magnetic field and thus energy $\mathcal{E}$, we empirically find that a cutoff of
\begin{equation}
    r_{\mathrm{cut}} = 50 a_0,
\end{equation}
with $a_0$ the Bohr radius results in good agreement between the wavefunctions. The insets of Fig.~\ref{fig:energy}d compare the pseudopotential, ab-initio and non-interacting wavefunctions for the $\text{BR}_1$ branch at three characteristic points at $\mathcal{E} \approx \frac{5}{2}\hbar\omega$ (Fig.~\ref{fig:energy}e, close to non-interacting oscillator state $\ket{1}_{\mathrm{rel}}$), $\mathcal{E} \approx \frac{7}{2}\hbar\omega$ (Fig.~\ref{fig:energy}f, at unitarity) and $\mathcal{E} \approx \frac{9}{2}\hbar\omega$ (Fig.~\ref{fig:energy}g, close to non-interacting oscillator state $\ket{3}_{\mathrm{rel}}$). The cutoff works well for all energies $\mathcal{E}$ and all trap depths explored in the experiment. We note that the chosen $r_\mathrm{cut}$ is close to the near-resonant effective range $R_0$ and also close to the characteristic van der Waals length $r_{\mathrm{VdW}} = 65 a_0$ for ground-state $^{40}$K atoms. Similar techniques were applied in Ref.~\cite{chipresonance}, and found analogous short-range thresholds.

Calculations that only require the long-range component of the interacting wavefunction use the pseudopotential result $\psi_{\mathrm{int}}(r)$ with the cutoff $r_{\mathrm{cut}} = 50 a_0$ unless otherwise specified. This wavefunction represents the $bb$ channel only; in the main text we denote the interacting state as $\ket{T,\uparrow\uparrow}$ for notational simplicity, but generically it is a superposition of multiple spin channels.

%%%%%%%%%%%%%%%%%%%%%%
\bigskip \noindent{\bf Anharmonicity corrections.}
%%%%%%%%%%%%%%%%%%%%%%
The spectra probed in the experiment are modified due to anharmonicity induced by the lattice potential. The full potential is given by
\begin{equation}
\hat{H}_{\mathrm{trap}} = V_\mathrm{L} \sum_{\alpha=1,2}\sum_{i=x,y,z} \sin^2\left(\frac{\pi \hat{x}_{i,\alpha}}{a}\right),
\end{equation}
where $V_\mathrm{L}$ is the lattice depth and $a$ the lattice spacing. We write this potential in spherical relative and center-of-mass coordinates $(r,\theta_r,\phi_r)$ and $(R,\theta_R,\phi_R)$. We assume that the atoms are at close distances ($r \ll a$) and near the center of the lattice site ($R \ll a$). Under this assumption we Taylor expand $\hat{H}_{\mathrm{trap}}$ to fourth order in $r$ and $R$,
\begin{equation}
\hat{H}_{\mathrm{trap}} = \hat{H}_{\mathrm{trap}}^{(2)} + \hat{H}_{\mathrm{trap}}^{(4)} + \mathcal{O}(r^6, r^4 R^2, r^2 R^4, R^6).
\end{equation}
The second-order term is given by
\begin{equation}
    \hat{H}_{\mathrm{trap}}^{(2)} = V_{\mathrm{L}} \frac{\pi^2}{a^2} \left(\frac{1}{2}r^2 + 2R^2\right),
\end{equation}
and gives the effective trapping frequency
\begin{equation}
    \omega = \sqrt{\frac{2 V_\mathrm{L} \pi^2}{a^2 m}},
\end{equation}
before any anharmonic corrections are included. The fourth-order term is
\begin{equation}
\begin{aligned}
\hat{H}_{\mathrm{trap}}^{(4)} = -V \frac{\pi^4}{a^4} [&\alpha(\theta_r,\phi_r) r^4 + \beta(\theta_r,\phi_r,\theta_R,\phi_R) r^2 R^2  \\
&+ \gamma(\theta_R,\phi_R)R^4 ],
\end{aligned}
\end{equation}
where the angular functions are given by

\begin{equation}
\begin{aligned}
\alpha(\theta_r,&\phi_r) =\frac{4\cos^4\!{\theta_r} + \sin^4\!{\theta_r}[3+\cos(4\phi_r)]}{96},\\
\beta(\theta_r,&\phi_r,\theta_R,\phi_R) = \cos^2\!{\theta_r} \cos^2\!{\theta_R} \\ & + \frac{1}{2}\sin^2\!{\theta_R} \sin^2\!{\theta_R} [1+\cos(2\phi_r)\cos(2\phi_R)],\\
\gamma(\theta_R,&\phi_R) =\frac{4\cos^4\!{\theta_R} + \sin^4\!{\theta_R}\,[3+\cos(4\phi_R)]}{6}.
\end{aligned}
\end{equation}
The lowest-order anharmonic correction to the spectrum can be estimated as the expectation value of the perturbing term $\hat{H}_{\mathrm{trap}}^{(4)}$.

The full spatial wavefunction of the initial state in our spectroscopy including angular components and center-of-mass motion can be written, assuming a Cartesian excitation along the $\hat{x}$ direction (although the results are identical for any single excitation in the $\hat{x}$-$\hat{y}$ plane), as
\begin{equation}
\begin{aligned}
\psi_{\mathrm{initial}} =& \psi_{\mathrm{rel}}^{(n_{\mathrm{rel}}=1)}(r) \frac{Y_{1,1}(\theta_r,\phi_r) + Y_{1,-1}(\theta_r,\phi_r)}{\sqrt{2}} \\ 
&\times \; \psi_{\mathrm{CM}}^{(n_{\mathrm{CM}}=0)}(R) Y_{0,0} (\theta_R, \phi_R),
\end{aligned}
\end{equation}
where the radial wavefunctions $\psi_{\mathrm{rel}}^{(n_{\mathrm{rel}}=1)}(r)$ and $\psi_{\mathrm{CM}}^{(n_{\mathrm{CM}}=0)}(R)$ are given explicitly by
\begin{equation}
\begin{aligned}
    \psi_{\mathrm{rel}}^{(n_{\mathrm{rel}}=1)}(r) &= \sqrt{\frac{8}{3 \sqrt{\pi} a_{\mathrm{ho}}^3}} \frac{r}{a_{\mathrm{ho}}} e^{-\frac{r^2}{2 a_{\mathrm{ho}}^2}},\\
    \psi_{\mathrm{CM}}^{(n_{\mathrm{CM}}=0)}(R) &= \sqrt{\frac{32}{\sqrt{\pi} a_{\mathrm{ho}}^3}} e^{-\frac{2R^2}{ a_{\mathrm{ho}}^2}}.
\end{aligned}
\end{equation}

Since the interactions only depend on $r$, if we assume that coupling between relative and center-of-mass motion induced by the lattice anharmonicity is negligible, the $p$-wave interacting state will have the same wavefunction for all coordinates except $r$:
\begin{equation}
\begin{aligned}
    \psi_{\mathrm{final}} = \psi_{\mathrm{int}}(r) & \frac{Y_{1,1}(\theta_r,\phi_r) + Y_{1,-1}(\theta_r,\phi_r)}{\sqrt{2}} \\ \times \; & \psi_{\mathrm{CM}}^{(n_{\mathrm{CM}}=0)}(R) Y_{0,0} (\theta_R, \phi_R),
\end{aligned}
\end{equation}
where $\psi_{\mathrm{int}}(r)$ is the interacting wavefunction from Eq.~\eqref{eq_ExactWavefunctionPseudopotential}. The correction to the energy values obtained by spectroscopy is then given by the difference of the anharmonic corrections between these two states,
\begin{equation}
\begin{aligned} 
   \Delta \mathcal{E}_{\mathrm{anharmonic}} = & \bra{\psi_{\mathrm{final}}} \hat{H}_{\mathrm{trap}}^{(4)} \ket{\psi_{\mathrm{final}}} \\ &- \bra{\psi_{\mathrm{initial}}} \hat{H}_{\mathrm{trap}}^{(4)} \ket{\psi_{\mathrm{initial}}}.
\end{aligned}
\end{equation}
This yields
\begin{widetext}
\begin{equation}
\begin{aligned}
\Delta \mathcal{E}_{\mathrm{anharmonic}} &= -V\frac{\pi^4}{a^4} \left[\frac{1}{40}\int_{0}^{\infty}dr \text{ } r^6 [\psi_{\mathrm{int}}(r)]^2 +\frac{1}{4} a_{\mathrm{ho}}^2 \int_{0}^{\infty}dr \text{ } r^4 [\psi_{\mathrm{int}}(r)]^2 - \frac{17}{8} a_{\mathrm{ho}}^4\right]\\
&=-E_R \left[\frac{1}{10a_{\mathrm{ho}}^4}\int_{0}^{\infty}dr \text{ } r^6 [\psi_{\mathrm{int}}(r)]^2 +\frac{1}{2a_{\mathrm{ho}}^2} \int_{0}^{\infty}dr \text{ } r^4 [\psi_{\mathrm{int}}(r)]^2 - \frac{17}{8} \right].
\end{aligned}
\end{equation}
\end{widetext}

%%%%%%%%%%%%%%%%%%%%%%
\bigskip \noindent{\bf Rabi oscillations.}
%%%%%%%%%%%%%%%%%%%%%%
The coherent oscillation of atomic population between the spin states $\downarrow,\uparrow$ under the RF drive has contributions from single-particle and pair dynamics. The single-particle Hamiltonian is given by $\hat{H}_{\mathrm{sing}} = \frac12 {\hbar \Omega_1} (e^{-i  2\pi f_{\mathrm{RF}} t} \hat{\sigma}^{+}+h.c.) + \frac12 { h f_{\mathrm{ZS}}} \hat{\sigma}^{z}$, where $\hat{\sigma}^{+}$, $\hat{\sigma}^{z}$ are standard Pauli operators, $\Omega_1$ is the single-photon Rabi frequency, $f_{\mathrm{RF}}$ is the RF frequency and $h f_{\mathrm{ZS}}$ is the Zeeman splitting. The Rabi frequency is assumed to be independent of $f_{\mathrm{RF}}$ as the variation of the laser wavevector $2\pi / f_{\mathrm{RF}}$ is negligible across the length of a lattice site. We apply the usual unitary transformation $\hat{U}_{\mathrm{sing}} =\exp [-i t (2 \pi f_\mathrm{RF}) \hat{\sigma}^{z}/2 ]$ which yields $\hat{H}_{\mathrm{sing}} \to \hat{U}_{\mathrm{sing}}^{\dagger}\hat{H}_{\mathrm{sing}}\hat{U}_{\mathrm{sing}} + i \hbar\hat{U}_{\mathrm{sing}}\frac{ d}{dt}\hat{U}^{\dagger}_{\mathrm{sing}}$, then set the RF frequency to $\delta/2\pi=f_{\mathrm{RF}} - f_{\mathrm{ZS}}$ with $\delta$ the detuning. The resulting Hamiltonian matrix is
\begin{equation}
    \hat{H}_{\mathrm{sing}} = \frac{\hbar}{2} \left(\begin{array}{cc} \delta & \Omega_1 \\ \Omega_1 & -\delta \end{array}\right)
\end{equation}
in the basis of single-atom spin states $ \{\ket{\downarrow},\ket{\uparrow}\}$. For an initial state $\ket{\downarrow}$ the number of excited atoms $N^{\uparrow}_{\mathrm{sing}} = \ket{\uparrow}\bra{\uparrow}$ is given by
\begin{equation}
    N^{\uparrow}_{\mathrm{sing}}(t) = \frac{\Omega_1^2}{\Omega_1^2 +\delta^2} \sin^2 \left(\frac{\sqrt{\Omega_1^2 + \delta^2}}{2} t\right).
\end{equation}
The generalised singlon Rabi frequency is thus set by $\widetilde{\Omega}_{1} = \sqrt{\Omega_1^2 + \delta^2}$.

For a pair, we write the Hamiltonian as 
\be \hat{H}_{\mathrm{doub}}= \frac{\hbar \Omega_1}{2}\sum_{\alpha}\left(e^{-i 2\pi f_{\mathrm{RF}} t}\hat{\sigma}_{\alpha}^{+}+h.c.\right)+\frac{h f_{\mathrm{ZS}}}{2} \sum_{\alpha}\hat{\sigma}_{\alpha}^{z}\,, \ee 
where $\hat{\sigma}_{\alpha}^{+}$ and $\hat{\sigma}_{\alpha}^{z}$ are Pauli operators acting on atom $\alpha \in \{1,2\}$. We make the same unitary transformation $\hat{U}_{\mathrm{doub}}=\exp(-i t 2\pi  \frac{f_{\mathrm{RF}}}{2} \sum_{\alpha}\hat{\sigma}^{z}_{\alpha})$ yielding $\hat{H}_{\mathrm{doub}} \to \hat{U}_{\mathrm{doub}}^{\dagger}\hat{H}_{\mathrm{doub}}\hat{U}_{\mathrm{doub}}+i \hbar \hat{U}_{\mathrm{doub}}\frac{ d}{dt}\hat{U}^{\dagger}_{\mathrm{doub}}$. 
The result is $\hat{H}_{\mathrm{doub}} = \frac12 {\hbar\Omega_1} \sum_{\alpha}\hat{\sigma}_{\alpha}^{x} - \frac12 {\hbar\delta} \sum_{\alpha}\hat{\sigma}_{\alpha}^{z}$. We next rewrite the Hamiltonian in the basis of the singlet and triplet states $\{\ket{T,\downarrow\downarrow},\ket{T,\uparrow\downarrow},\ket{T,\uparrow\uparrow},\ket{S,\uparrow\downarrow}\}$. The latter singlet state decouples and can be dropped from the calculations. The Hamiltonian for the remaining triplet states, including the interaction term $U_p^{(1)}\ket{T,\uparrow\uparrow}\bra{T,\uparrow\uparrow}$ for the branch probed in the experiment, is
\begin{equation}
    \hat{H}_{\mathrm{doub}} = \frac{\hbar}{2} \left(\begin{array}{ccc} 2\delta & {\sqrt{2}}{\Omega_1} & 0 \\ {\sqrt{2}}{\Omega_1} & 0 & {\sqrt{2}}{\Omega_1}\eta \\
    0 & {\sqrt{2}}{\Omega_1}\eta & U_p^{(1)}\!/\hbar-2\delta\end{array}\right).
\end{equation}
The basis for this model including the relative motional excitations is $\{\ket{T,\downarrow\downarrow} \ket{1}_{\mathrm{rel}}$, $ \ket{T,\uparrow\downarrow} \ket{1}_{\mathrm{rel}}$, $  \ket{T,\uparrow\uparrow} \ket{\psi_{\mathrm{int}}}_{\mathrm{rel}}\}$, where $\ket{\psi_{\mathrm{int}}}_{\mathrm{rel}}$ is the spatial wavefunction of the interacting state in the relative coordinate $r$. We write this basis as $\{\ket{\downarrow\downarrow}, \ket{\uparrow\downarrow}, \ket{\uparrow\uparrow}\}$ hereafter for brevity. Note that the single-photon Rabi frequencies in this Hamiltonian are larger than the corresponding singlon frequency by a factor of $\sqrt{2}$ due to constructive interference. The parameter $\eta$ is the spatial wavefunction overlap in the relative coordinate between the interacting and non-interacting states,
\begin{equation}
\label{eq_EtaDefinitionSupplement}
    \eta = {}_{\mathrm{rel}}\langle\psi_{\mathrm{int}} |1\rangle_{\mathrm{rel}} = \int_0^{\infty}dr\> r^2 \psi_{\mathrm{int}}^{*}(r) \psi_{\mathrm{rel}}^{(n_{\mathrm{rel}}=1)}(r).
\end{equation}

The two-photon detuning is chosen to match the interaction energy $U_p^{(1)}$ of the $p$-wave interacting state in the $\text{BR}_1$ branch, in order to maximise the amplitude of the Rabi oscillations between $\ket{\downarrow\downarrow}$ and $\ket{\uparrow\uparrow}$. In this case ($\hbar \delta = U_{p}^{(1)}/2$), the spin-triplet hamiltonian becomes
\begin{equation}
   \hat{H}_{\mathrm{doub}} = \frac{\hbar}{2} \left(\begin{array}{ccc} 0 & {\sqrt{2}}{\Omega_1} & 0 \\ {\sqrt{2}}{\Omega_1} & -{U_{p}^{(1)}}/{\hbar} & {\sqrt{2}}{\Omega_1}\eta \\
    0 & {\sqrt{2}}{\Omega_1}\eta & 0\end{array}\right) + \text{Const}.
\end{equation}

When starting from $\ket{\downarrow\downarrow}$, the time-dependent excited state fraction $N^{\uparrow}_{\mathrm{doub}} = \ket{\uparrow\downarrow}\bra{\uparrow\downarrow} + 2 \ket{\uparrow\uparrow}\bra{\uparrow\uparrow}$ for this Hamiltonian is given by
\begin{widetext}
\begin{equation}
\begin{aligned}
    N^{\uparrow}_{\mathrm{doub}}(t) &= \frac{\eta^2 \Omega_1^2}{(1+\eta^2) \nu^2} \left[\frac{1-\eta^2}{\eta^2}\sin^2 \left(\frac{\nu}{2} t\right) + \frac{8 \nu}{2\nu-U_{p}^{(1)}\!/\hbar} \sin^2 \left(\frac{2\nu-U_{p}^{(1)}\!/\hbar}{8} t\right) + \frac{8 \nu}{2\nu+U_{p}^{(1)}\!/\hbar} \sin^2 \left(\frac{2\nu+U_{p}^{(1)}\!/\hbar}{8} t\right)\right],\\
    \nu &= \frac{1}{2}\sqrt{(U_{p}^{(1)}\!/\hbar)^2 + 8(1 + \eta^2)\Omega_1^2}\>.
\end{aligned}
\end{equation}
\end{widetext}
In the limit of weak single-photon Rabi coupling $U_{p}^{(1)} \gg \hbar \Omega_1$ (for positive $U_{p}^{(1)} > 0$) this expression simplifies to
\begin{equation}
\begin{aligned}
    N^{\uparrow}_{\mathrm{doublon}}(t) =& \frac{8 \eta^2}{(1+\eta^2)^2} \sin^2\!\left( \frac{2\nu-U_{p}^{(1)}\!/\hbar}{8} t \right) \\ &+ \mathcal{O}\! \left( \frac{\hbar^2 \Omega_1^2}{(U_{p}^{(1)})^2}\right).
\end{aligned}
\end{equation}
The oscillations of the pairs are characterised by a single frequency
\begin{equation}
\begin{aligned}
  \widetilde{\Omega}_2 =& \frac{2 \nu - U_{p}^{(1)}\!/\hbar}{4} \\
  =& \frac{\sqrt{(U_{p}^{(1)}\!/\hbar)^2 + 8(1+\eta^2)\Omega_1^2}- U_{p}^{(1)}\!/\hbar}{4}.
\end{aligned}
\end{equation}
This is the pair Rabi oscillation frequency observed in the oscillation of magnetisation. Since all quantities apart from the wavefunction overlap $\eta$ are known or measured independently, the experimentally measured $\widetilde{\Omega}_2$ directly translates into an extracted $\eta$, which results in the experimental points shown in Fig.~\ref{fig:rabi}e.

The theoretical prediction obtains $\eta$ by directly computing a wavefunction overlap using either the pseudopotential wavefunction [via Eq.~\eqref{eq_EtaDefinitionSupplement}] or the ab-initio wavefunctions,
\begin{equation}
    \eta \approx \int_0^{\infty}dr\> r^2 \psi_{bb}^{*}(r) \psi_{\mathrm{rel}}^{(n_{\mathrm{rel}}=1)}(r).
\end{equation}
For the ab-initio case we only use the $bb$ channel because it is the only spin channel with non-negligible wavefunction amplitude at long range $r \gtrsim a_{\mathrm{ho}}$. Other spin channels would have additional prefactors in the single-photon coupling matrix element caused by Clebsch-Gordan coefficients, but their overlap with the oscillator state is negligible (even if their overall probability is not).

%%%%%%%%%%%%%%%%%%%%%
\bigskip \noindent{\bf Lifetime of the interacting state.}
%%%%%%%%%%%%%%%%%%%%%
The lifetime of the pairs depends on the short range state fraction $\chi$. To estimate $\chi$ we use the ab-initio calculations together with a coarse-grained approach that treats all short-range wavefunction population, independent of spin channel, as a lossy state fraction responsible for decay. Figure~\ref{fig:lifetime}c shows the total wavefunction probability density summed over all channels from the ab-initio calculation $r^2|\psi(r)|^2 = r^2\sum_{\sigma}|\psi_{\sigma}(r)|^2$. For all magnetic field values there is a clear distinction between a short range component with amplitude at $r < a_{\mathrm{ho}}$ and a long-range component with amplitude at $r \gtrsim a_{\mathrm{ho}}$, with negligible population between the two regimes. We estimate the lossy fraction by establishing an empirical threshold,
\begin{equation}
    \chi = \sum_{\sigma} \int_0^{r_{\mathrm{short}}} dr \>r^2 |\psi_{\sigma}(r)|^2.
\end{equation}
The threshold separation $r_{\mathrm{short}}$ is chosen to capture all of the wavefunction norm of the short range part only. A value of $r_{\mathrm{short}} = 325 a_0$ captures the short-range fraction for the field values and trap depths probed in the lifetime measurements, allowing for the good agreement with the measurements. We emphasise however that this approach is not strongly sensitive to the particular choice of $r_{\mathrm{short}}$: as the threshold separation $r_{\mathrm{short}}$ is increased to capture most of the short-range probability, the theory prediction saturates to the solid curves shown in Figs.~\ref{fig:lifetime}e-f. 

This coarse-grained approach treating all spin channels $\sigma$ on equal footing works well because at short range, the spatial wavefunctions $\psi_{\sigma}(r)$ have approximately the same functional form (up to overall prefactors) since the dipolar interactions are dominant in that regime. The coupling of each spin channel wavefunction to the lossy dimer state is then approximately the same. Furthermore, while the dimer state itself is not fully equivalent to the free-space state for which the $\tau_d = 3.4$ ms was calculated in Ref.~\cite{chipresonance} due to the trap, we can still use that lifetime because the additional energy imparted by the trap is negligible compared to the short-range interaction scales.

One can also use this approach to obtain $\chi$ from the pseudopotential wavefunction,
\begin{equation}
    \chi = \int_{0}^{r_{\mathrm{short}}}dr\> r^2 |\psi_{\mathrm{int}}(r)|^2,
\end{equation}
which leads to the dashed curves in Figs.~\ref{fig:lifetime}e-f showing similar agreement for the same threshold.

\end{document}